\documentclass[twocolumn,showpacs,preprintnumbers,amsmath,amssymb]{revtex4}

\usepackage{graphicx}
\usepackage{dcolumn}
\usepackage{bm}
\usepackage{multirow}

\newcommand{\snn}{\ensuremath{\sqrt{s_{_{NN}}}}}
\newcommand{\pim}{\ensuremath{\pi^{-}\pi^{-}}}
\newcommand{\pip}{\ensuremath{\pi^{+}\pi^{+}}}
\newcommand{\dedx}{\ensuremath{\langle dE/dx \rangle}}
\newcommand{\npart}{\ensuremath{\langle N_{part} \rangle}}
\newcommand{\cc}[1]{\multicolumn{1}{c}{#1}}
\newcommand{\ykp}{\ensuremath{y_{_{\it YKP}}}}
\newcommand{\qinv}{\ensuremath{\mathbf q_{inv}}}

\newcolumntype{z}{D{.}{.}{13}}%
\newcolumntype{y}{D{.}{.}{2.15}}%
\newcolumntype{x}{D{.}{.}{20}}%

\begin{document}


\title{Transverse momentum and rapidity dependence of HBT
correlations\\in Au+Au collisions at $\snn =$ 62.4 and 200 GeV} 

\author{
B.B.Back$^1$,
M.D.Baker$^2$,
M.Ballintijn$^4$,
D.S.Barton$^2$,
R.R.Betts$^6$,
A.A.Bickley$^7$,
R.Bindel$^7$,
A.Budzanowski$^3$,
W.Busza$^4$,
A.Carroll$^2$,
Z.Chai$^2$,
M.P.Decowski$^4$,
E.Garc\'{\i}a$^6$,
T.Gburek$^3$,
N.George$^{1,2}$,
K.Gulbrandsen$^4$,
S.Gushue$^2$,
C.Halliwell$^6$,
J.Hamblen$^8$,
M.Hauer$^2$,
G.A.Heintzelman$^2$,
C.Henderson$^4$,
D.J.Hofman$^6$,
R.S.Hollis$^6$,
R.Ho\l y\'{n}ski$^3$,
B.Holzman$^2$,
A.Iordanova$^6$,
E.Johnson$^8$,
J.L.Kane$^4$,
J.Katzy$^{4,6}$,
N.Khan$^8$,
W.Kucewicz$^6$,
P.Kulinich$^4$,
C.M.Kuo$^5$,
W.T.Lin$^5$,
S.Manly$^8$,
D.McLeod$^6$,
A.C.Mignerey$^7$,
R.Nouicer$^6$,
A.Olszewski$^3$,
R.Pak$^2$,
I.C.Park$^8$,
H.Pernegger$^4$,
C.Reed$^4$,
L.P.Remsberg$^2$,
M.Reuter$^6$,
C.Roland$^4$,
G.Roland$^4$,
L.Rosenberg$^4$,
J.Sagerer$^6$,
P.Sarin$^4$,
P.Sawicki$^3$,
H.Seals$^2$,
I.Sedykh$^2$,
W.Skulski$^8$,
C.E.Smith$^6$,
M.A.Stankiewicz$^2$,
P.Steinberg$^2$,
G.S.F.Stephans$^4$,
A.Sukhanov$^2$,
J.-L.Tang$^5$,
M.B.Tonjes$^7$,
A.Trzupek$^3$,
C.Vale$^4$,
G.J.van~Nieuwenhuizen$^4$,
S.S.Vaurynovich$^4$,
R.Verdier$^4$,
G.I.Veres$^4$,
E.Wenger$^4$,
F.L.H.Wolfs$^8$,
B.Wosiek$^3$,
K.Wo\'{z}niak$^3$,
A.H.Wuosmaa$^1$,
B.Wys\l ouch$^4$\\
(PHOBOS Collaboration)\\
\vspace{3mm}
\small
$^1$~Argonne National Laboratory, Argonne, IL 60439-4843, USA\\
$^2$~Brookhaven National Laboratory, Upton, NY 11973-5000, USA\\
$^3$~Institute of Nuclear Physics PAN, Krak\'{o}w, Poland\\
$^4$~Massachusetts Institute of Technology, Cambridge, MA 02139-4307, USA\\
$^5$~National Central University, Chung-Li, Taiwan\\
$^6$~University of Illinois at Chicago, Chicago, IL 60607-7059, USA\\
$^7$~University of Maryland, College Park, MD 20742, USA\\
$^8$~University of Rochester, Rochester, NY 14627, USA\\
}

\date{\today}

\begin{abstract}
Two-particle correlations of identical charged pion pairs from Au+Au collisions at 
$\snn = 62.4$ and 200 GeV were measured by the PHOBOS experiment at RHIC. Data for
the 15\% most central events were analyzed with Bertsch-Pratt and 
Yano-Koonin-Podgoretskii parameterizations using pairs with
rapidities of  $0.4 < y_{\pi\pi} < 1.3$ and transverse momenta $0.1 < k_T < 1.4$ GeV/c.
The Bertsch-Pratt radii $R_{o}$ and $R_{\ell}$ decrease as a function of pair transverse momentum, while
$R_{s}$ is consistent with a weaker dependence.  
$R_{o}$ and $R_{s}$ are independent of collision energy, while
$R_{\ell}$ shows a slight increase.
The source rapidity $\ykp$ scales roughly with
the pair rapidity $y_{\pi\pi}$, indicating strong dynamical correlations.
\end{abstract}

\pacs{25.75.-q,25.75.Dw,25.75.Gz}

\maketitle

Recent experimental results from all four experiments at the Relativistic Heavy
Ion Collider (RHIC) have concluded that Au+Au collisions at the top RHIC energy 
($\snn = 200$~GeV) have produced an extremely hot and dense state of matter
\cite{phob_da,star_da,phen_da,bram_da}. This matter may have degrees of freedom that 
are purely hadronic (``hadronic gas''), purely partonic (``Quark-Gluon Plasma'', QGP), 
or an admixture of both.  Identical-particle correlation measurements (Hanbury-Brown 
and Twiss, HBT) yield valuable information on the size, shape,
duration, and spatiotemporal evolution of the emission source.  
Because the dynamics of a hadron gas and a QGP are na\"{\i}vely expected to be quite 
different, HBT may allow us to discriminate between these three scenarios
\cite{pratt84,pratt86}.  

Experimentally, the correlation function $C({\mathbf q})$ is defined as
\begin{equation}
\centering
C({\mathbf q}) = \frac{P({\mathbf p}_1, {\mathbf p}_2)}{P({\mathbf p}_1)P({\mathbf p}_2)}
\end{equation}
where ${\mathbf p}_1$ and ${\mathbf p}_2$ are the particle four-momenta, $P({\mathbf p}_1, {\mathbf p}_2)$ is 
the probability of a pair being measured 
with relative four-momentum $\mathbf{q} = \mathbf{p}_1 - \mathbf{p}_2$, and
$P(\mathbf{p}_1)$ and $P(\mathbf{p}_2)$ are the single-particle probabilities. 
The numerator is determined directly from data, while the denominator is
constructed using a standard event-mixing technique.

$C({\mathbf{q}})$ can be fit to the Bertsch-Pratt parameterization of
a Gaussian source in three dimensions 
\cite{bertschparam,pratt86,crossterm},
\begin{equation}
\label{eq:osl+cross}
C({\mathbf{q}})  =
1 + \lambda e^{-(       
        q_{o}^2 R_{o}^2 + 
        q_{s}^2 R_{s}^2 +
        q_{\ell}^2 R_{\ell}^2 + 
        2 q_{o} q_{\ell} R_{o\ell}^2)}
\end{equation}
where $q_{\ell}$ is the component of $\mathbf{q}$ along the beam
direction; $q_{o}$ is the component along the pair transverse momentum
$\vec{k}_T = \frac{1}{2}(\vec{p}_{_{T1}} + \vec{p}_{_{T2}})$; and $q_{s}$ is the component
orthogonal to the other two. The $q_{o}q_{\ell}$ cross-term vanishes only for
symmetric collisions with
acceptances centered around midrapidity.  The $\lambda$ parameter
represents the correlation strength and is expected to be unity for a
completely incoherent source. According to the definitions of 
$q_{o}$ and $q_{s}$, $R_{o}$ probes a mixture of the
spatial and temporal extent of the source, while $R_{s}$ measures only
the spatial component. In the special case of a boost-invariant,
transparent, azimuthally symmetric source, the ratio $R_{o}/R_{s}$ may
be a good indicator of the duration of the emission of particles from the source. 
Predictions for this quantity from hydrodynamic and transport models varied by over an 
order of magnitude \cite{gyulassy96,soff}, but mostly focused on values between 
1.5--2.0, while the first results from RHIC at $\snn = 130$ GeV
indicated a value near unity \cite{star130,phenix130}.   
A more recent three-dimensional
hydrodynamic calculation (including opacity and tranverse flow)
\cite{morita} also was unable to reproduce the experimentally
measured values for $R_{o}/R_{s}$.
The results at $\snn = 130$ GeV also were consistent with a monotonic increase
of $R_{\ell}$ from AGS energies ($\snn \simeq$ 2--5 GeV), while 
$R_{o}$ and $R_{s}$ remained roughly constant.
It should be noted that it is 
possible to fit the experimental data at RHIC to a variety of \emph{ans\" atze}
\cite{budalund,blastwave,renk}, 
but the interpretation remains
an open question.  The wealth of HBT data over a large number of variables 
and their 
experimentally-determined systematic dependencies continue to create new 
challenges for theorists and add yet more pieces to the ``HBT puzzle''.

The data reported here for Au+Au collisions at $\snn = 62.4$ and 200 GeV were 
collected using the PHOBOS two-arm magnetic spectrometer during RHIC Run IV 
(2004) and Run II (2001), respectively. Details of the setup
have been described previously \cite{phob_NIM}.  The spectrometer
arms are each equipped with 16 layers of silicon sensors, providing
charged particle reconstruction both outside and inside a 2 T magnetic
field.  The two symmetric arms and
frequent magnetic field polarity reversals allowed for a number of
independent cross-checks in the two-particle correlation measurement.
A two layer silicon detector covering $| \eta | < 0.9$
and 25\% of the azimuthal angle provided additional information on the
position of the primary collision vertex.

The primary event trigger was provided by two sets of 16 scintillator
paddle counters, which covered a pseudorapidity range $3 < | \eta | <
4.5$. 
More information on PHOBOS event selection and
centrality determination can be found in \cite{phobos1,phobos2}.

The details of the track reconstruction algorithm can be found in
\cite{phob_pbarp130,phob_spectra200}.  Events with a reconstructed
primary vertex position between -12 cm $< z_{vtx} <$ 10 cm along the
beam direction and -0.1 $< x_{vtx} <$ 0.2 cm and -0.05 $< y_{vtx} <$ 0.2 cm
along the transverse directions
were selected to optimize vertex-finding precision,
track reconstruction efficiency, and momentum resolution.  Only
particles which traversed the entire spectrometer were used in the
analysis.  A $3\sigma$ cut on the distance of closest approach of each
reconstructed track with respect to the primary vertex ($dca_{vtx} <
0.35$ cm) was used to reject background particles from decays and
secondary interactions.
\begin{figure}
\includegraphics[width=8cm]{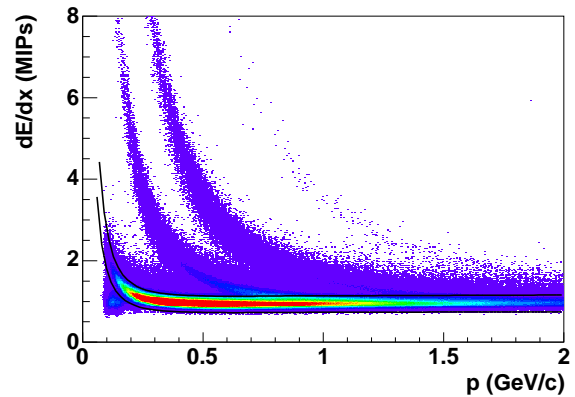} 
\caption{ \label{fig:Selected_pions} Distribution of the truncated
average energy loss $\dedx$ as a function of reconstructed particle
momentum.  The region between the solid black lines corresponds to the $\pm 3\sigma$ cut used
to select pions for this analysis.}
\end{figure}
The final track selection was based on the $\chi^{2}$ probability of a
full track fit, taking into account multiple scattering and energy
loss.  The momentum resolution is $\Delta p/p \sim 1 \%$ after all
cuts.  Particle identification was based on the truncated mean of the
specific ionization loss in the silicon detectors.  To identify pions,
a cut of three RMS deviations away from the expected mean value of the
specific ionization $\dedx$ for pions was applied.  This is shown as the
region between the solid black lines
in Fig. \ref{fig:Selected_pions}.  Possible contamination
from other particle species was studied using HIJING 1.35~\cite{HIJING}
and a GEANT 3.21~\cite{GEANT} simulation of the full detector, applying
the same reconstruction procedures to Monte Carlo events and data.
The contamination from 
$\textrm{K}^+ \textrm{K}^+$, $\textrm{K}^- \textrm{K}^-$, pp, $\overline{\textrm{p}}\,\overline{\textrm{p}}$,
$\textrm{e}^+ \textrm{e}^+$, and $\textrm{e}^- \textrm{e}^-$ pairs is estimated to be less than 1\%;
non-identical pairs are estimated to contribute less than 10\% throughout the entire
$k_{T}$ range.  To reject ghost pairs, only one shared hit out of
six in the
weak-field region and two shared hits out of five in the strong-field region were
allowed per pair.  A two-particle acceptance cut was applied to both
data and background; the criterion for pair acceptance was defined by
$\Delta \phi + 2 \Delta \theta > 0.05$ rad, where $\Delta\phi$ and
$\Delta\theta$ are the relative pair separation in azimuthal and polar
angle, respectively.  
In this analysis, events in the most central 15\% of the cross section
were selected. HIJING was used to relate the fraction
of the cross section to $\npart$, the mean number of participating
nucleons \cite{phobos1}.  The average number of participants for these
events was estimated to be $\npart = 310$ (303) at $\snn = 200$ (62.4) GeV.
About 7.3 (3.3) million $\pip$ and 5.5 (3.0) million
$\pim$ pairs for $\snn = 200$ (62.4) GeV survive all cuts including  
centrality selection.

The event-mixed background is constructed by combining single tracks from 
randomly selected different events into fake pairs.
Events from widely spaced collision vertices may yield
a background that cannot reproduce the same phase space as the actual data.
This effect is particularly prominent for a small-acceptance
detector such as PHOBOS.   In this analysis, an ``event class'' is 
defined solely by the vertex resolution of the detector.
Events were defined as belonging to the same event class if
they had vertices within $0.025$ cm along the beam axis and $0.05$ cm along
the vertical and horizontal axes.
Background events are only constructed from tracks mixed from the same event
class.

Systematic errors (90\% C.L.) were determined by changing cuts in two-particle acceptance
and single-track azimuthal acceptance, using different random seeds for 
mixed-event background generation, as well as varying the event class definition
to create background events from pairs with narrower and broader vertex ranges.
Additionally, these studies were performed for each individual spectrometer arm
and the differences included in the systematic uncertainties.

Because the background is constructed from tracks belonging to
different events, it does not {\it a priori} include multiparticle
correlations (apart from a residual effect at the
few percent level, which has been corrected 
following \cite{zajc84,cianciolo_thesis}). 
In order to study the HBT
correlation, it is necessary to apply a weight to account for the
Coulomb effect.  The Coulomb correction can be expressed solely as a 
function of the invariant relative 4-momentum $\qinv$,
\begin{equation}
F_{R_{inv}}(\qinv) = \frac{F_c(\qinv)}{F_{pl}(\qinv)} = \frac{\int d\vec{r}\,
|\psi_c(\vec{r})|^2 S (\vec{r})} {\int d\vec{r}\,
|\psi_{pl}(\vec{r})|^2 S(\vec{r})}
\end{equation}
where $S(\vec{r})$ is the distribution of the relative separation of the particle pairs
at emission, $R_{inv}$ is the radius parameter conjugate to $\qinv$,
and $\psi_c$ and $\psi_{pl}$ are the Coulomb and plane wave-functions,
respectively.  A closed-form approximation and numerical interpolation
for this relation was derived in \cite{859_hbt} for $\lambda = 1$.

For variable $\lambda$ \cite{holzman04},  
\begin{equation}
F_{R_{inv}}(\qinv, \lambda) = \frac{(1-\lambda) + \lambda(1 + e^{-{\qinv^2R_{inv}^2})} F_{R_{inv}}(\qinv)}
                       {1 + \lambda e^{-{\qinv}^2R_{inv}^2} }
\end{equation}

This prescription was derived simultaneously and 
is nearly equivalent to the corrections applied by
the CERES \cite{ceres}, STAR \cite{star200}, and PHENIX experiments \cite{phenix200};
our results showed no significant
change using either correction method.  The method is applied
iteratively, successively fitting distributions of the correlation
function $C(\qinv)$ and applying the fit values $\lambda$ and $R_{inv}$ to a new
$S(\vec{r})$.  Typically 2 or 3 iterations are sufficient for
convergence.

The three-dimensional correlation functions were fit by Eq. 
(\ref{eq:osl+cross}) using MINUIT and the log-likelihood method. 
Table \ref{table1} shows the results of the fit for 
both $\pip$ and $\pim$. The data were analyzed 
in the longitudinal co-moving system (LCMS) frame.
\begin{figure}
\includegraphics[width=8cm]{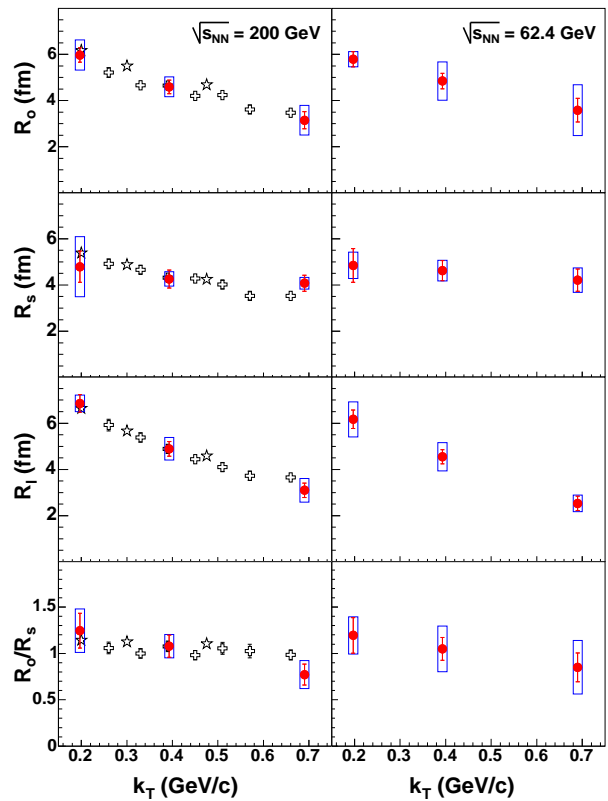} 
\caption{ \label{fig:pim_comp}
Bertsch-Pratt parameters $R_{o}$, $R_{s}$, $R_{\ell}$ and the 
ratio $R_{o}/R_{s}$ 
for $\pim$ pairs at $\snn = 200$ GeV (left panel) and 62.4 GeV (right panel) 
as a function of $\langle k_{T} \rangle$.   
For comparison, data from STAR \cite{star200} (open stars) and 
PHENIX \cite{phenix200} (open crosses) are presented at 200 GeV.  
PHOBOS systematic errors (90\% C.L.) are shown as boxes; systematic errors from
STAR and PHENIX are not shown.}
\end{figure}
In Fig. \ref{fig:pim_comp}, the Bertsch-Pratt parameters and the ratio 
$R_{o}/R_{s}$ are presented as a function of $k_{T}$ for $\pim$
pairs at $\snn = 62.4$ GeV and $200$ GeV.  The 200 GeV data are compared 
to data from STAR \cite{star200} and PHENIX \cite{phenix200}.
Not shown are values for $\lambda$, which is roughly $0.5$,  
and the cross-term $R_{ol}^2$, which is consistent
with zero for all $k_T$ bins.
The small vertical acceptance of the PHOBOS detector roughly translates into a 
small acceptance in $q_{s}$, which is responsible for the large
statistical and systematic uncertainties on the value of $R_{s}$ at low $k_T$.   
Within errors, the 
values of $R_{s}$ are consistent with a weak dependence on $k_{T}$, 
while the values of $R_{o}$ and $R_{\ell}$ decrease rapidly with increasing 
$k_{T}$.  Consequently, the ratio $R_{o}/R_{s}$  
decreases from $1.24 \pm 0.19$ to $0.77 \pm 0.11$ for $\langle k_{T} \rangle$ from 0.2 to 0.7 GeV/c.
It should be noted that the STAR and PHENIX data are measured at midrapidity while the 
PHOBOS data are measured
at $\langle y_{\pi\pi} \rangle  = 0.9$; additionally, STAR, PHOBOS, and PHENIX data are from the
top 10\%, 15\%, and 30\% of the cross-section respectively.

Despite the very different experimental
acceptances, the data at 200 GeV agree remarkably well.  The trends at $\snn = 62.4$ GeV are qualitatively and
quantitatively similar to those at 200 GeV, except that $R_{\ell}$ may be slightly smaller.

\begin{figure}
\includegraphics[width=8cm]{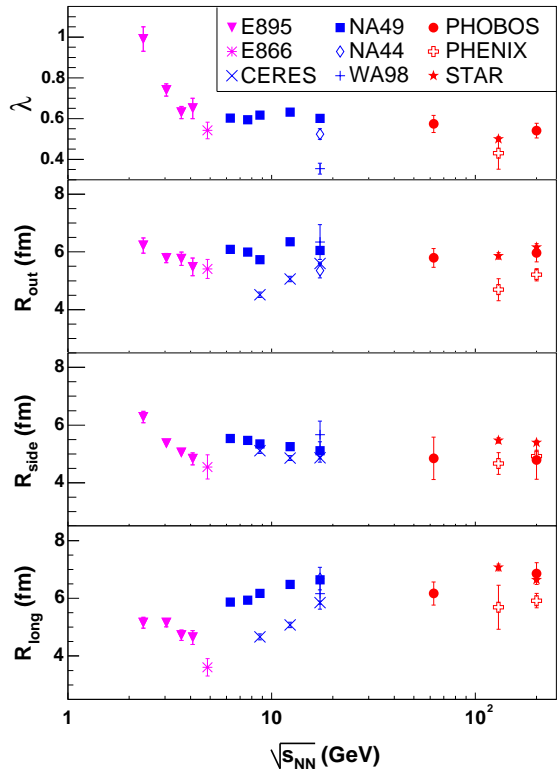}
\caption{\label{fig:excit}
Bertsch-Pratt parameters $\lambda$, $R_{s}$, $R_{o}$, and $R_{\ell}$ as a function of $\snn$ for $\pim$ pairs.
The presented data are near midrapidity and represent comparable $k_T$ bins from each 
experiment
 \cite{ceres,e895,859_hbt,na44,na49,wa98,star130,star200,phenix130,phenix200}.  PHOBOS data
are represented by solid circles.  Systematic errors are not shown.
}
\end{figure}

The Bertsch-Pratt parameters $\lambda$, $R_{o}$, $R_{s}$, and $R_{\ell}$ are 
presented in Fig. \ref{fig:excit} as a function of 
center of mass energy from $\snn = 2$ to 200 GeV.
The data are $\pim$ pairs near midrapidity for comparable $k_T$ bins from 
nine different experiments.  The data at $\snn = 62.4$ GeV begin to fill the large gap 
between the top
SPS energy and the 130 GeV RHIC data.  Although there is some
disagreement among experiments, the data do not appear to exhibit any sharp 
discontinuities, but smoothly vary as a function of collision energy.
The correlation function was also fit to the Yano-Koonin-Podgoretskii (YKP)
parameterization \cite{yanokoonin,podgoretskii} 
\begin{equation}
\label{eq:ykp}
C({\mathbf{q}})  =
1 + \lambda e^{-(       
        q_{\perp}^2 R_{\perp}^2 + 
        \gamma^2(q_{\|}-\beta q_{\tau})^2 R_{\|}^2+
        \gamma^2(q_{\tau}-\beta q_{\|})^2 R_{\tau}^2         
        )}
\end{equation}
where $\beta$ is the longitudinal velocity of the source and $\gamma =
1/\sqrt{1-\beta^2}$, $q_{\perp}$ and $q_{\|}$ are the relative 3-momentum
difference projected in the transverse and longitudinal directions,
respectively, and $q_{\tau}$ is the relative difference in energy.  Employing
this coordinate system provides the advantange of factorizing the velocity
and duration of the source from the spatial parameters.  
In addition to fits in the LCMS frame, data were also fit in the laboratory frame as a 
cross-check and yielded consistent results.

\begin{figure}
\includegraphics[width=8cm]{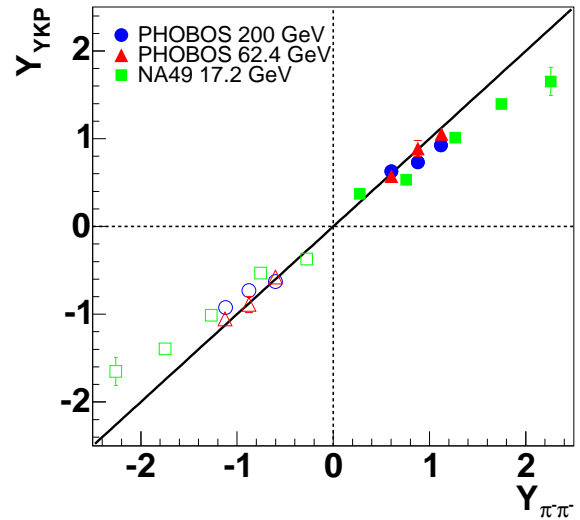} 
\caption{ \label{fig:pho_yykp}
Dependence of the source rapidity $\ykp$ vs. the pair rapidity 
$y_{\pi\pi}$ for PHOBOS at 200 GeV (circles), PHOBOS at 62.4 GeV (triangles),
and NA49 (negatively charged hadrons) at 17.2 GeV \cite{na49ykp} (squares). 
Open symbols represent data reflected about midrapidity.
}
\end{figure}

Table \ref{table2} shows the results of Yano-Koonin-Podgoretskii
fits for $\pip$ and $\pim$ at both energies.
In Fig. \ref{fig:pho_yykp}, the value of the source
rapidity $\ykp$ (as extracted from $\beta$ in Eq. \ref{eq:ykp}) is
plotted as a function of pair rapidity for $\pim$ pairs with
$0.1 < k_T < 1.4$~GeV/c. Open circles indicate the results reflected
about midrapidity.
The data from NA49 \cite{na49ykp} at 
$\snn = 17.2$~GeV is also 
plotted; however, it should be noted the NA49
data presented here cover only $0.1 < k_T < 0.2$~GeV/c.
The source rapidity scales with the rapidity of the pair, indicating the presence of strong
position-momentum correlations. A static source would exhibit no 
correlation and would correspond to a horizontal line ($\ykp = 0$).
A source with strong dynamical correlations would correspond to a
straight line along $\ykp = y_{\pi\pi}$.  The data are consistent
with the latter scenario: particles emitted at a given rapidity were produced
by a source moving collectively at the same rapidity.
\begin{table*}
\begin{center}
\caption{Bertsch-Pratt fit parameters for $\pip$ and
$\pim$ in the rapidity range $0.4 < y_{\pi\pi} <1.3$ for the 15\% most central
Au+Au collisions. Momenta are in GeV/c and radii are in fm.  Errors are
statistical ($1\sigma$) followed by systematic (90\% C.L.)}
\label{table1}
\newcommand{\m}{\hphantom{$-$}}
\renewcommand{\arraystretch}{1.5} 
\begin{tabular}{cccyyyzzzz}
\hline
\cc{Species} & \cc{$\snn$} &
\cc{$\langle k_{T} \rangle$} & \cc{$\lambda$} & \cc{$R_{o}$} & \cc{$R_{s}$} & \cc{$R_{l}$} &
\cc{$R_{ol}^{2}$ (fm$^2$)} \\ 
\hline
\multirow{3}{*}{$\pip$}&
& 0.198&
0.620 \pm 0.046 \pm 0.075 &
5.65 \pm 0.33 \pm 0.44 &
5.11 \pm 0.69 \pm 0.61 &
6.70 \pm 0.44 \pm 0.63 &
1.9 \pm 2.9 \pm 2.6 \\ &
62.4 GeV
& 0.392&
0.660 \pm 0.055 \pm 0.069 &
4.72 \pm 0.34 \pm 0.48 &
4.74 \pm 0.45 \pm 0.42 &
4.53 \pm 0.31 \pm 0.31 &
0.85 \pm 1.7 \pm 2.2 \\ &
& 0.680&
0.413 \pm 0.073 \pm 0.083 &
2.70 \pm 0.60 \pm 0.66 &
3.94 \pm 0.46 \pm 0.67 &
2.21 \pm 0.52 \pm 1.3 &
0.26 \pm 0.91 \pm 0.91 \\
\hline
\multirow{3}{*}{$\pim$}&
& 0.198&
0.574 \pm 0.042 \pm 0.057 &
5.79 \pm 0.33 \pm 0.33 &
4.85 \pm 0.73 \pm 0.57 &
6.17 \pm 0.40 \pm 0.76 &
2.2 \pm 2.9 \pm 2.9 \\ &
62.4 GeV
& 0.391&
0.688 \pm 0.059 \pm 0.070 &
4.84 \pm 0.34 \pm 0.83 &
4.62 \pm 0.44 \pm 0.45 &
4.55 \pm 0.31 \pm 0.61 &
0.33 \pm 1.9 \pm 1.4 \\ &
& 0.673&
0.532 \pm 0.075 \pm 0.13 &
3.58 \pm 0.51 \pm 1.1 &
4.21 \pm 0.48 \pm 0.53 &
2.53 \pm 0.32 \pm 0.36 &
-0.32 \pm 1.2 \pm 2.0 \\
\hline
\hline
\multirow{3}{*}{$\pip$}&
& 0.198&
0.493 \pm 0.028 \pm 0.041 &
5.48 \pm 0.27 \pm 0.36 &
4.09 \pm 0.63 \pm 0.66 &
6.75 \pm 0.33 \pm 0.26 &
4.9 \pm 2.3 \pm 5.7 \\ &
200 GeV
& 0.393&
0.582 \pm 0.041 \pm 0.086 &
5.05 \pm 0.28 \pm 0.40 &
4.64 \pm 0.39 \pm 0.40 &
4.75 \pm 0.25 \pm 0.52 &
0.96 \pm 1.7 \pm 2.0 \\ &
& 0.685&
0.444 \pm 0.049 \pm 0.082 &
3.31 \pm 0.32 \pm 0.52 &
3.96 \pm 0.36 \pm 0.20 &
2.98 \pm 0.29 \pm 0.37 &
0.25 \pm 1.1 \pm 1.5 \\
\hline
\multirow{3}{*}{$\pim$}&
& 0.198&
0.541 \pm 0.036 \pm 0.048 &
5.97 \pm 0.31 \pm 0.65 &
4.79 \pm 0.67 \pm 1.3 &
6.86 \pm 0.38 \pm 0.36 &
5.7 \pm 3.0 \pm 2.2 \\ &
200 GeV
& 0.393&
0.559 \pm 0.044 \pm 0.054 &
4.59 \pm 0.30 \pm 0.43 &
4.26 \pm 0.39 \pm 0.31 &
4.89 \pm 0.32 \pm 0.49 &
1.1 \pm 1.9 \pm 3.0 \\ &
& 0.690&
0.469 \pm 0.055 \pm 0.060 &
3.15 \pm 0.37 \pm 0.64 &
4.08 \pm 0.35 \pm 0.25 &
3.10 \pm 0.31 \pm 0.51 &
0.93 \pm 1.1 \pm 2.1 \\
\hline
\end{tabular}
\end{center}
\end{table*}
\begin{table*}
\begin{center}
\caption{Yano-Koonin-Podgoretskii fit parameters for $\pip$
and $\pim$ in the $k_T$ range $0.1 < k_{T} < 1.4$ GeV/c for the 15\% most central
Au+Au collisions. Radii are in fm. Errors are
statistical ($1\sigma$) followed by systematic (90\% C.L.)}
\label{table2}
\renewcommand{\arraystretch}{1.5} 
\begin{tabular}{cccyyyzzzz}
\hline
\cc{Species} & \cc{$\snn$} &
\cc{$\langle y_{\pi\pi} \rangle$} & \cc{$\lambda$} & \cc{$R_{\perp}$} & \cc{$R_{\|}$} & \cc{$R_{\tau}$} & \cc{$\beta$}  \\
\hline
\multirow{3}{*}{$\pip$}&
& 0.602&
0.669 \pm 0.057 \pm 0.074 &
5.05 \pm 0.33 \pm 0.30 &
5.75 \pm 0.45 \pm 0.70 &
0.0017 \pm 2.0 \pm 1.4 &
-0.162 \pm 0.086 \pm 0.11 \\ &
62.4 GeV
& 0.879&
0.512 \pm 0.051 \pm 0.10 &
4.21 \pm 0.62 \pm 0.52 &
4.88 \pm 0.44 \pm 0.48 &
3.7 \pm 1.3 \pm 3.0 &
0.119 \pm 0.095 \pm 0.091 \\ &
& 1.124&
0.506 \pm 0.046 \pm 0.062 &
3.96 \pm 0.42 \pm 0.41 &
5.20 \pm 0.47 \pm 0.58 &
3.0 \pm 1.0 \pm 2.0 &
-0.0738 \pm 0.080 \pm 0.082 \\
\hline
\multirow{3}{*}{$\pim$}&
& 0.602&
0.596 \pm 0.055 \pm 0.067 &
4.40 \pm 0.70 \pm 1.0 &
5.07 \pm 0.44 \pm 0.41 &
4.3 \pm 1.2 \pm 1.3 &
-0.0266 \pm 0.079 \pm 0.098 \\ &
62.4 GeV
& 0.878&
0.539 \pm 0.054 \pm 0.080 &
4.78 \pm 0.64 \pm 0.69 &
5.10 \pm 0.48 \pm 0.82 &
0.75 \pm 6.4 \pm 2.6 &
0.0105 \pm 0.13 \pm 0.23 \\ &
& 1.124&
0.525 \pm 0.044 \pm 0.040 &
4.09 \pm 0.44 \pm 0.47 &
4.84 \pm 0.35 \pm 0.25 &
3.3 \pm 0.99 \pm 2.0 &
-0.0687 \pm 0.075 \pm 0.048 \\
\hline
\hline
\multirow{3}{*}{$\pip$}&
& 0.602&
0.507 \pm 0.037 \pm 0.056 &
4.83 \pm 0.53 \pm 0.48 &
5.54 \pm 0.37 \pm 0.36 &
2.3 \pm 1.7 \pm 2.3 &
-0.0967 \pm 0.077 \pm 0.087 \\ &
200 GeV
& 0.878&
0.441 \pm 0.034 \pm 0.088 &
4.36 \pm 0.53 \pm 0.73 &
5.37 \pm 0.37 \pm 0.82 &
1.3 \pm 2.9 \pm 2.2 &
-0.130 \pm 0.091 \pm 0.23 \\ &
& 1.119&
0.454 \pm 0.031 \pm 0.032 &
3.89 \pm 0.37 \pm 0.59 &
5.03 \pm 0.31 \pm 0.43 &
3.6 \pm 0.70 \pm 1.0 &
-0.115 \pm 0.058 \pm 0.069 \\
\hline
\multirow{3}{*}{$\pim$}&
& 0.602&
0.492 \pm 0.040 \pm 0.042 &
3.83 \pm 0.55 \pm 0.44 &
5.43 \pm 0.38 \pm 0.45 &
4.4 \pm 0.96 \pm 1.3 &
0.0270 \pm 0.070 \pm 0.056 \\ &
200 GeV
& 0.877&
0.485 \pm 0.042 \pm 0.079 &
4.50 \pm 0.50 \pm 0.70 &
5.23 \pm 0.40 \pm 0.71 &
2.3 \pm 1.6 \pm 1.7 &
-0.146 \pm 0.090 \pm 0.14 \\ &
& 1.122&
0.469 \pm 0.040 \pm 0.044 &
4.16 \pm 0.44 \pm 0.31 &
5.72 \pm 0.43 \pm 0.57 &
2.2 \pm 1.3 \pm 0.90 &
-0.193 \pm 0.080 \pm 0.11 \\
\hline
\end{tabular}
\end{center}
\end{table*}
In conclusion, we have measured HBT parameters in Au+Au collisions with the PHOBOS detector
using both the Bertsch-Pratt and Yano-Koonin-Podgoretskii parameterizations
at energies of $\snn = 62.4$ GeV and 200 GeV.  The ratio $R_{o}/R_{s}$ 
does not show significant deviation from unity over the entire $k_{T}$ range.
The data at 200 GeV show a nice agreement between three RHIC experiments with different
acceptances, centrality measures, and rapidity, which implies that the source parameters are fairly
insensitive to these variables at this energy.  
Additionally, the Bertsch-Pratt
parameters evolve smoothly as a function of $\snn$ over nearly two orders of magnitude. 

This work was partially supported by U.S. DOE grants 
DE-AC02-98CH10886,
DE-FG02-93ER40802, 
DE-FC02-94ER40818,  
DE-FG02-94ER40865, 
DE-FG02-99ER41099, and
W-31-109-ENG-38, by U.S. 
NSF grants 9603486, 
0072204,            
and 0245011,        
by Polish KBN grant 1-PO3B-062-27(2004-2007), and
by NSC of Taiwan Contract NSC 89-2112-M-008-024.

\end{document}